\documentclass[prd,onecolumn,nofootinbib,showkeys]{revtex4}
\usepackage{bm,amsmath,amssymb}
\newcommand\gothfamily{\usefont{U}{ygoth}{m}{n}}
\DeclareTextFontCommand{\textgoth}{\gothfamily}

\begin{document}

\title{COSMOLOGICAL CONSTANT FROM QUARKS AND TORSION}

\author{{\bf Nikodem J. Pop{\l}awski}}

\affiliation{Department of Physics, Indiana University, Swain Hall West, 727 East Third Street, Bloomington, Indiana 47405, USA}
\email{nikodem.poplawski@gmail.com}

\noindent
{\em Annalen der Physik} ({\em Berlin})\\
Vol. {\bf 523}, No. 4 (2011) pp. 291--295\\
\copyright\,WILEY-VCH Verlag GmbH \& Co. KGaA, Weinheim
\vspace{0.4in}

\begin{abstract}
We present a simple and natural way to derive the observed small, positive cosmological constant from the gravitational interaction of condensing fermions.
In the Riemann-Cartan spacetime, torsion gives rise to the axial-axial four-fermion interaction term in the Dirac Lagrangian for spinor fields.
We show that this nonlinear term acts like a cosmological constant if these fields have a nonzero vacuum expectation value.
For quark fields in QCD, such a torsion-induced cosmological constant is positive and its energy scale is only about 8 times larger than the observed value.
Adding leptons to this picture could lower this scale to the observed value.
\end{abstract}


\maketitle

A positive cosmological constant in the Einstein equations for the gravitational field is the simplest form of dark energy, a yet unexplained energy that causes the observed current acceleration of the Universe \cite{obs}.
Quantum field theory predicts that the corresponding vacuum energy density is on the order of $m^4_{\textrm{Pl}}$, where $m_{\textrm{Pl}}$ is the reduced Planck mass, which is about 120 orders of magnitude larger than the measured value $\rho_\Lambda=(2.3\,\mbox{meV})^4$.
This cosmological-constant problem is thus the worst problem of fine-tuning in physics.
Zel'dovich argued, using dimensional analysis, that the cosmological vacuum energy density should be on the order of $\rho_\Lambda\sim m^6/m^2_{\textrm{Pl}}$, where $m$ is the mass scale of elementary particles \cite{Zel,ED}.
However, some theoretical arguments have been used to show that the cosmological constant must vanish \cite{Li}.
It is possible that the huge value of a cosmological constant from the zero-point energy of vacuum may be cancelled out by an effective cosmological term arising from spinning fluids in the Riemann-Cartan spacetime \cite{dSS} or reduced through some dynamical processes \cite{Ab}.
It is also possible that the observed osmological constant is simply another fundamental constant of Nature \cite{BR}.

A model of a cosmological constant caused by the vacuum expectation value in quantum chromodynamics (QCD) through QCD trace anomaly from gluonic and quark condensates gives $\rho_\Lambda\sim H\lambda^3_{\textrm{QCD}}$ \cite{Sch}, where $H$ is the Hubble parameter and $\lambda_{\textrm{QCD}}\approx 200\,\mbox{MeV}$ is the QCD scale parameter of the SU(3) gauge coupling constant \cite{Sh}.
If a cosmological constant is caused by the vacuum energy density from the gluon condensate of QCD then $\rho_\Lambda\sim\lambda^6_{\textrm{QCD}}/m^2_{\textrm{Pl}}$ \cite{KV1}, which resembles the formula of Zel'dovich \cite{Zel}.
Another QCD-derived model of a cosmological constant gives $\rho_\Lambda\sim Hm_q\langle q\bar{q}\rangle/m_{\eta'}$, where $\langle q\bar{q}\rangle$ is the chiral quark condensate \cite{UZ}.
A cosmological constant may be also caused by the vacuum energy density from the electroweak phase transition, giving $\rho_\Lambda\sim E^8_{\textrm{EW}}/m^4_{\textrm{Pl}}$, where $E_{\textrm{EW}}$ is the energy scale of this transition \cite{KV2}.
The cosmic acceleration could also arise from a Bardeen-Cooper-Schrieffer condensate of fermions in the presence of torsion, which forms in the early Universe \cite{Alex}, or from dark spinors \cite{BB}.

In this paper, we present a simple and natural way to derive the small, positive cosmological constant from fermionic condensates and the Einstein-Cartan-Sciama-Kibble theory of gravity with torsion.
Such a constant arises from a vacuum expectation value of the Dirac-Heisenberg-Ivanenko-Hehl-Datta four-fermion interaction term in the Lagrangian for quark (and lepton) fields.
Thus the cosmological constant may simply originate from particle physics and relativistic gravity with spin.

The Einstein-Cartan-Sciama-Kibble (ECSK) theory of gravity \cite{KS} naturally extends Einstein's general relativity to include matter with intrinsic half-integer spin, which produces torsion, providing a more complete account of local gauge invariance with respect to the Poincar\'{e} group \cite{HD,Hehl}.
The Riemann spacetime of general relativity is generalized to the Riemann-Cartan spacetime with torsion.
The ECSK gravity is a viable theory, which differs significantly from general relativity only at densities of matter much larger than the density of nuclear matter.
Torsion may also prevent the formation of singularities from matter with spin \cite{avert,ISS}, averaged as a spin fluid \cite{WR}, and appears to introduce an effective ultraviolet cutoff in quantum field theory for fermions \cite{Niko}.
Moreover, torsion fields may cause the current cosmic acceleration \cite{Nes}.

In the Riemann-Cartan spacetime, the Dirac Lagrangian density is given by $\mathfrak{L}=\frac{i\sqrt{-g}}{2}(\bar{\psi}\gamma^i\psi_{;i}-\bar{\psi}_{;i}\gamma^i\psi)-m\sqrt{-g}\bar{\psi}\psi$, where the semicolon denotes a full covariant derivative with respect to the affine connection.
Varying $\mathfrak{L}$ with respect to spinor fields gives the Dirac equation with a full covariant derivative.
Varying the total Lagrangian density $-\frac{R\sqrt{-g}}{2\kappa}+\mathfrak{L}$ with respect to the torsion tensor gives the relation between the torsion and the Dirac spin density which is quadratic in spinor fields \cite{HD,Hehl}.
Substituting this relation to the Dirac equation gives the nonlinear (cubic) Dirac-Heisenberg-Ivanenko-Hehl-Datta equation for $\psi$ (in the units in which $\hbar=c=1$, $\kappa=m^{-2}_{\textrm{Pl}}$) \cite{HD,Hehl}:
\begin{equation}
i\gamma^k\psi_{:k}=m\psi-\frac{3\kappa}{8}(\bar{\psi}\gamma_k\gamma^5\psi)\gamma^k\gamma^5\psi,
\label{HI}
\end{equation}
where the colon denotes a covariant derivative with respect to the Christoffel symbols.
This equation and its adjoint conjugate can also be obtained directly by varying, respectively over $\bar{\psi}$ and $\psi$, the following effective Lagrangian density \cite{HD}:
\begin{equation}
\mathfrak{L}_e=\frac{i\sqrt{-g}}{2}(\bar{\psi}\gamma^i\psi_{:i}-\bar{\psi}_{:i}\gamma^i\psi)-m\sqrt{-g}\bar{\psi}\psi+\frac{3\kappa\sqrt{-g}}{16}(\bar{\psi}\gamma_k\gamma^5\psi)(\bar{\psi}\gamma^k\gamma^5\psi),
\label{Lagr}
\end{equation}
without varying it with respect to the torsion.
The corresponding effective energy-momentum tensor $T_{ik}=\frac{2}{\sqrt{-g}}\frac{\delta\mathfrak{L}_e}{\delta g^{ik}}$ is, using the identity $\frac{\delta\gamma^j}{\delta g^{ik}}=\frac{1}{2}\delta^j_{(i}\gamma_{k)}$ (which results from the definition of the Dirac matrices, $\gamma^{(i}\gamma^{k)}=g^{ik}I$), given by:
\begin{equation}
T_{ik}=\frac{i}{2}(\bar{\psi}\delta^j_{(i}\gamma_{k)}\psi_{:j}-\bar{\psi}_{:j}\delta^j_{(i}\gamma_{k)}\psi)-\frac{i}{2}(\bar{\psi}\gamma^j\psi_{:j}-\bar{\psi}_{:j}\gamma^j\psi)g_{ik}+m\bar{\psi}\psi g_{ik}-\frac{3\kappa}{16}(\bar{\psi}\gamma_j\gamma^5\psi)(\bar{\psi}\gamma^j\gamma^5\psi)g_{ik}.
\label{emt}
\end{equation}
Substituting (\ref{HI}) into (\ref{emt}) gives
\begin{equation}
T_{ik}=\frac{i}{2}(\bar{\psi}\delta^j_{(i}\gamma_{k)}\psi_{:j}-\bar{\psi}_{:j}\delta^j_{(i}\gamma_{k)}\psi)+\frac{3\kappa}{16}(\bar{\psi}\gamma_j\gamma^5\psi)(\bar{\psi}\gamma^j\gamma^5\psi)g_{ik}.
\label{ten}
\end{equation}
The first term on the right of (\ref{ten}) is the energy-momentum tensor for a Dirac field without torsion while the second term corresponds to an effective cosmological constant \cite{dSS,ISS,Ker},
\begin{equation}
\Lambda=\frac{3\kappa^2}{16}(\bar{\psi}\gamma_j\gamma^5\psi)(\bar{\psi}\gamma^j\gamma^5\psi),
\end{equation}
or a vacuum energy density,
\begin{equation}
\rho_\Lambda=\frac{3\kappa}{16}(\bar{\psi}\gamma_j\gamma^5\psi)(\bar{\psi}\gamma^j\gamma^5\psi).
\end{equation}

Such a torsion-induced cosmological constant depends on spinor fields, so it is not constant in time (it is constant in space at cosmological scales in a homogeneous and isotropic universe).
However, if these fields can form a condensate then the vacuum expectation value of $\Lambda$ will behave like a real cosmological constant.
Quark fields in QCD form a condensate with the nonzero vacuum expectation value for $\bar{\psi}\psi$,
\begin{equation}
\langle0|\bar{\psi}\psi|0\rangle\approx-(230\,\mbox{MeV})^3\sim-\lambda_{\textrm{QCD}}^3.
\label{vev}
\end{equation}
In the Shifman-Vainshtein-Zakharov vacuum-state-dominance approximation, the matrix element $\langle 0|\bar{\psi}\Gamma_1\psi\bar{\psi}\Gamma_2\psi|0\rangle$, where $\Gamma_1$ and $\Gamma_2$ are any matrices from the set $\{I,\gamma^i,\gamma^{[i}\gamma^{k]},\gamma^5,\gamma^5\gamma^i\}$, can be reduced to the square of $\langle0|\bar{\psi}\psi|0\rangle$ \cite{SVZ}:
\begin{equation}
\langle0|\bar{\psi}\Gamma_1\psi\bar{\psi}\Gamma_2\psi|0\rangle=\frac{1}{12^2}\Bigl((tr\Gamma_1\cdot tr\Gamma_2)-tr(\Gamma_1\Gamma_2)\Bigr)\times(\langle0|\bar{\psi}\psi|0\rangle)^2.
\label{vsd}
\end{equation}
For quark fields, we have $\Gamma_1=\gamma_i\gamma^5 t^a$ and $\Gamma_2=\gamma^i\gamma^5 t^a$, where $t^a$ are the Gell-Mann matrices acting in the color space and normalized by the condition $tr(t^a t^b)=2\delta^{ab}$.
Thus we obtain
\begin{equation}
\langle0|(\bar{\psi}\gamma_j\gamma^5 t^a\psi)(\bar{\psi}\gamma^j\gamma^5 t^a\psi)|0\rangle=\frac{16}{9}(\langle0|\bar{\psi}\psi|0\rangle)^2,
\end{equation}
which gives
\begin{equation}
\langle0|\rho_\Lambda|0\rangle=\frac{\kappa}{3}(\langle0|\bar{\psi}\psi|0\rangle)^2,
\label{cc}
\end{equation}
corresponding to a positive cosmological constant.
This formula resembles celebrated Zel'dovich's relation \cite{Zel}, with the mass scale of elementary particles $m$ corresponding to $(-\langle 0|\bar{\psi}\psi|0\rangle)^{1/3}$.
Combining this relation with the expression for $\rho_\Lambda$ in \cite{Sch} gives $Hm_{\textrm{Pl}}^2\sim\lambda_{\textrm{QCD}}^3$.
Interestingly, using a Lorentz-violating axial condensate instead of the QCD quark vacuum condensates leads to a very similar result \cite{BJ}.
Substituting (\ref{vev}) into (\ref{cc}) gives
\begin{equation}
\langle 0|\rho_\Lambda|0\rangle\approx(54\,\mbox{meV})^4.
\end{equation}
The value of the observed cosmological constant would agree with the torsion-induced cosmological constant presented here if $\langle 0|\bar{\psi}\psi|0\rangle$ were $\approx-(28\,\mbox{MeV})^3$, suggesting a contribution to $\Lambda$ from spinor fields with a lower (in magnitude) vacuum expectation value.
Such fields could correspond to neutrinos \cite{neu}.

The presented model combines the ECSK gravity, which is the simplest theory with torsion, and QCD.
It predicts a positive cosmological constant due to: the axial-axial form of the four-fermion interaction term in the Dirac Lagrangian (\ref{Lagr}), the vacuum-state-dominance formula for SU(3) (\ref{vsd}), and the nonzero vacuum expectation value for quantum fields (\ref{vev}).
The vector-vector form of a four-fermion interaction would give a negative cosmological constant, but this form does not result from the ECSK theory with minimally coupled fermions.
It is possible, however, to modify the form of the quartic term by adding to the Lagrangian density $-\frac{R\sqrt{-g}}{2\kappa}+\mathfrak{L}$ two terms: one proportional to $R_{ijkl}\epsilon^{ijkl}$, related to the Barbero-Immirzi parameter \cite{BI}, and another proportional to $\frac{\sqrt{-g}}{2}(\bar{\psi}\gamma^i\psi_{;i}+\bar{\psi}_{;i}\gamma^i\psi)$, measuring the nonminimal coupling of fermions to gravity in the presence of torsion \cite{Fre}.

Although the four-fermion interaction term in (\ref{Lagr}) seems to be nonrenormalizable, we emphasize that this term appears in the effective Lagrangian density $\mathfrak{L}_e$ in which only the metric tensor and spinor fields are dynamical variables.
The original Lagrangian density $\mathfrak{L}$, in which the torsion tensor is also a dynamical variable, is renormalizable.
We also note that the torsion may modify the concept of renormalization by providing an effective ultraviolet cutoff for fermions \cite{Niko}.
Another problem could be: what cancels much larger contributions to the vacuum energy density arising from quantum field theory?
It has been argued in \cite{BR}, however, that vacuum energy does not gravitate; only a shift in vacuum energy (vacuum expectation value of physical fields) produces a gravitational field.
Therefore extremely large contributions to the vacuum energy density from quantum field theory should not appear in the Einstein equations.
These issues need to be investigated further.

The torsion in the ECSK theory is minimally coupled to spinor fields.
Thus the only parameter in this simple model of the cosmological constant is the energy of the two-quark condensate (\ref{vev}).
This model gives a cosmological constant whose energy scale is only about $\frac{230}{28}\approx 8$ times larger than that corresponding to the observed cosmological constant.
Therefore it provides the simplest explanation for the sign (and, to some extent, magnitude) of the observed cosmological constant.
We expect that adding lepton condensates to this picture could lower the average $|\langle 0|\bar{\psi}\psi|0\rangle|$ such that the resulting torsion-induced cosmological constant would agree with its observed value.
The absolute value of $\langle 0|\bar{\psi}\psi|0\rangle$ could also be lowered by introducing the two parameters considered in \cite{Fre}.
We also emphasize that our model naturally derives Zel'dovich's formula \cite{Zel} for the cosmological constant from a fundamental theory (the ECSK gravity coupled to Dirac fields), indicating that the results of this work are not a numerical coincidence.

\section*{Acknowledgements}
The author would like to thank James Bjorken for very interesting and fruitful discussions on torsion and modified theories of gravity.

\end{document}